\documentclass[12pt]{iopart}

\usepackage{iopams}
\usepackage[dvips]{graphics,color}
\usepackage[pdftex]{graphicx}

\def\bra#1{\mathinner{\langle{#1}|}}
\def\ket#1{\mathinner{|{#1}\rangle}}

\begin{document}

\title{Noise-assisted energy transfer in quantum networks and light-harvesting complexes}

\author{A W Chin$^{1}$, A Datta$^{2,3}$, F Caruso$^{1}$, S F Huelga$^{1}$ and M B Plenio$^{1,2,3}$}

\address{$^1$ Institut f{\"u}r Theoretische Physik, Universit{\"a}t Ulm, D-89069, Ulm, Germany}
\address{$^2$ Institute for Mathematical Sciences, Imperial College London, 53 Exhibition Road, London SW7 2PG, UK}
 \address{$^3$ QOLS, The Blackett Laboratory, Imperial College London, Prince Consort Road, London SW7 2BW, UK}

\ead{alex.chin@uni-ulm.de}

\begin{abstract}
We provide physically intuitive mechanisms for the effect of noise
on excitation energy transfer (EET) in networks. Using these
mechanisms of dephasing-assisted transport (DAT) in a hybrid basis
of both excitons and sites, we develop a detailed picture of how noise
enables energy transfer with efficiencies well above $90\%$ across the
Fenna-Matthew-Olson (FMO) complex, a type of light harvesting molecule.
We demonstrate explicitly how noise alters the pathways of energy
transfer across the complex, suppressing ineffective pathways and
facilitating direct ones to the reaction centre. We explain that the fundamental
mechanisms underpinning DAT are expected to be robust with respect to
the considered noise model but show that the specific details of the exciton-phonon
coupling, which remain largely unknown in these type of complexes, and in
particular the impact of non-Markovian effects, results in variations of dynamical
features that should be amenable to experimental verification within current or
planned technology. A detailed understanding of DAT in natural compounds
should open up a new paradigm of `noise-engineering' by
which EET can be optimized in artificial light-harvesting
structures.
\end{abstract}
\submitto{\NJP}
\maketitle

\section{Introduction}\label{intro}
The early stages of natural photosynthesis are able to capture and transport incident light energy with nearly $100\%$ efficiency, and a clear understanding of these processes could be immensely valuable for optimizing the efficiency of artificial light-harvesting devices~\cite{sension07}. Ultrafast nonlinear spectroscopy has been used to probe energy transfer dynamics in the Fenna-Matthew-Olson (FMO) complex~\cite{brixner05,fleming07a,fleming07b}, a crucial part of the photosynthetic system of green sulphur bacteria. Recently, further experiments at physiological temperatures have been performed \cite{engel10}. The FMO complex is an example of a pigment-protein complex (PPC), a network through which electronic excitations on individual pigments can migrate via excitonic couplings. It functions as a type of molecular `wire' that funnels light energy captured in the chlorosome antennae to a reaction center (RC) where the energy is used to initiate chemical reactions~\cite{olson04}. These experiments have demonstrated the existence of strong quantum coherences between multiple pigments, and have shown that the highly efficient energy relaxation in this system proceeds via coherently delocalized exciton states~\cite{brixner05}. In addition, wave-like beating between these excitons has also been observed to persist on timescales $>550$ fs, a significant fraction of the typical transport time in FMO~\cite{fleming07a}.

These observations have generated considerable interest in understanding the possibly functional role of quantum coherence effects in the remarkably efficient excitation energy transfer (EET) in FMO and other PPCs. It was initially argued that the fast transfer rates may be attributed to the exploitation of quantum search algorithms by the quantum dynamics of the FMO complex \cite{sension07,fleming07a}. However, typical timescales for relaxation and dephasing in PPCs \cite{Mukamel08} suggest that any quantum entanglement will be short-ranged, and as it is generally accepted that efficient quantum computation requires long range entanglement~\cite{jozsa}, the application of the principles of standard quantum computation in this system are far from straightforward. As a result, the complete formulation of a possible link between quantum coherence and functionality is still an open problem \cite{caro,coherence1,coherence2,coherence3}.

Starting with the seminal work of ~\cite{forster48,redfield65,grover71,kenre74,pearl72,haken73}, EET has been studied within the chemical physics community for several decades, yet the topic remains timely \cite{cheng00,scholes03,adolphs06,ishizaki09} due to the recent development of new experimental methods and also novel numerical techniques which allow theoretical models of the, still unknown, system-environment couplings to be tested\cite{numerics}.
Methods and ideas from quantum information science have also recently started to provide a new and complementary perspective on EET dynamics. Theoretical investigations of the role of pure dephasing noise in EET have found that this noise has the ability to enhance both the rate and yield of EET when compared to perfectly quantum coherent evolution~\cite{aspuruguzik08a,plenio08}. These results challenge the traditional view in information processing that noise always degrades the efficiency of quantum processes, and demonstrates that controllable noise can even be considered as an additional engineering tool for tasks like excitation transport\cite{chp}.

In the exciton basis normally used in previous studies~\cite{cho05}, dephasing-assisted transport (DAT) is understood as resulting from noise-induced transitions between exciton eigenstates states which cause energetic relaxation of the excitations towards the RC. Though
sufficient to suggest the possible existence of DAT in systems in contact with an uncontrollable environment, this approach does not currently describe in a transparent way \emph{how} does DAT actually work in detail and how it might be controlled or used. Such an understanding is essential for the fabrication of future artificial systems in which unavoidable noise might to some extent be employed as a constructive element that could use DAT processes to optimize light-harvesting and transport \cite{gaab04,perdomo10}.

This article revises and extends the physically intuitive picture we presented in \cite{caro} of how DAT and more general noise-assisted transport operates and highlights the engineering potential of deliberately applying noise in quantum transport networks. Starting with an idealized and exactly soluble model of noisy network transport, we revisit the underlying mechanisms that lead to DAT and then proceed to introduce additional complexity into this model to reveal how these mechanisms work in conjunction. Using these insights we analyze simulations of EET in FMO using a hybrid basis that allows a novel and clear visualization of how these mechanisms operate in this system and also make quantitative predictions about how they can optimize the transport. In essence, it allows us to follow the evolution of the initial excitation across the molecular `wire' as it makes its way to the reaction centre. The exact nature of the hybrid basis is governed by the relative magnitudes of the site energies and their couplings to the neighbouring sites. In fact, it has been recently (and independently) used to explain the experimentally observed quantum coherences in the photosynthetic apparatus of cryptophyte algae at room temperature~\cite{cwwcbs10}. Finally, we expand upon previous descriptions of DAT in FMO to consider forms of non-markovian dephasing and show that despite the fundamental mechanisms supporting DAT are robust with respect to the considered noise model, observable dynamical differences do appear which should be amenable to experimental testing. These type of evidence would be most valuable to discern the exact nature of the exciton-phonon coupling in different types of photosynthetic complexes.

\section{The Network model}\label{network}
Following previous theoretical descriptions of
PPCs~\cite{cheng00,aspuruguzik08a,plenio08}, we consider the PPCs as networks composed of distinct sites, one of which receives a single initial excitation, while another is connected to the RC. The network of $N$ sites is described by the coherent hopping Hamiltonian
\begin{equation}
\label{ham}
H = \sum_{j=1}^N \hbar\omega_j \sigma_j^{+}\sigma_j^{-} + \sum_{j\neq l} \hbar v_{j,l} (\sigma_j^{-}\sigma_{l}^{+} + \sigma_j^{+}\sigma_{l}^{-}),
\end{equation}
where $\sigma_j^{+}= \ket{j}\bra{0}$ and $\sigma_j^{-}=\ket{0}\bra{j}$ are raising and lowering operators for site $j$, the state $|j\rangle$ denotes one excitation in site $j$ and $|0\rangle$ is the zero exciton state. The local site energies are $\hbar\omega_j$, and $v_{j,l}$ are the coherent tunnelling amplitude between the sites $j$ and $l$. We do not assume any particular form for the microscopic coupling that generates these tunneling amplitudes and we therefore treat them as free parameters when considering abstract networks as we do in section \ref{fcn}. For the case of the FMO complex we will use the published Hamiltonian for \textit{P. aestuarii} taken from Ref.\cite{adolphs06}, where the hopping parameters are fixed by the geometry and dipolar-structure of the site interactions. We will further assume that the system is susceptible simultaneously to two distinct types of noise: A radiative decay process that transfers the excitation energy in site $j$ to the environment (with rate $\Gamma_j$) and a pure dephasing process (with rate $\gamma_j$) that destroys the phase coherence of any superposition of localized excitations. The dynamics of the network's density matrix is modelled by a Markovian master equation of the form
\begin{equation}
\dot{\rho}(t)=-i[H,\rho(t)]+{\cal L}_{rad}(\rho(t))+{\cal L}_{deph}(\rho(t)),\end{equation}
where the local radiative and pure dephasing terms are described respectively by Lindblad super-operators ${\cal L}_{deph}$ and ${\cal L}_{rad}$,
\begin{eqnarray}
{\cal L}_{deph}(\rho) &=& \sum_{j=1}^{N} \gamma_j[-\{\sigma_j^{+}\sigma_j^{-},\rho\} + 2 \sigma_j^{+}\sigma_j^{-}\rho \sigma_j^{+}\sigma_j^{-}],\\
 \label{dissipation}
{\cal L}_{rad}(\rho) &=& \sum_{j=1}^{N} \Gamma_j[-\{\sigma_j^{+}\sigma_j^{-},\rho\} + 2 \sigma_j^{-}\rho \sigma_j^{+} ],
\end{eqnarray}
where $\{A,B\}$ denotes an anticommutator. Formally, this approach is equivalent to the Haken-Strobl model at infinite temperature~\cite{haken73}, where pure dephasing is accounted for in terms of a classical, fluctuating field. The total excitation transfer is measured by the population transferred to the reaction center, modelled as the `sink' node, numbered $N+1$, which is populated by an irreversible decay process (with rate $\Gamma_{N+1}$) from a site $k$ of the network and described by a Lindblad operator ${\cal L}_{sink}(\rho)$
\begin{eqnarray}
{\cal L}_{sink}(\rho) &=& \Gamma_{N+1}\left[2\sigma_{N+1}^{+}\sigma_k^{-}\rho \sigma_k^{+}\sigma_{N+1}^{-} - \{\sigma_k^{+}\sigma_{N+1}^{-}\sigma_{N+1}^{+} \sigma_k^{-},\rho\} \right].
\end{eqnarray}
The model is completed by introducing the sink population
\begin{equation}
p_{sink}(t) =2\Gamma_{N+1}\int_{0}^t\rho_{kk}(t')\mathrm{d}t',\end{equation}
which will be used as the key measure of the transport efficiency.

\section{The fully-connected network}\label{fcn}
\begin{figure}[t]
\includegraphics[width=5in]{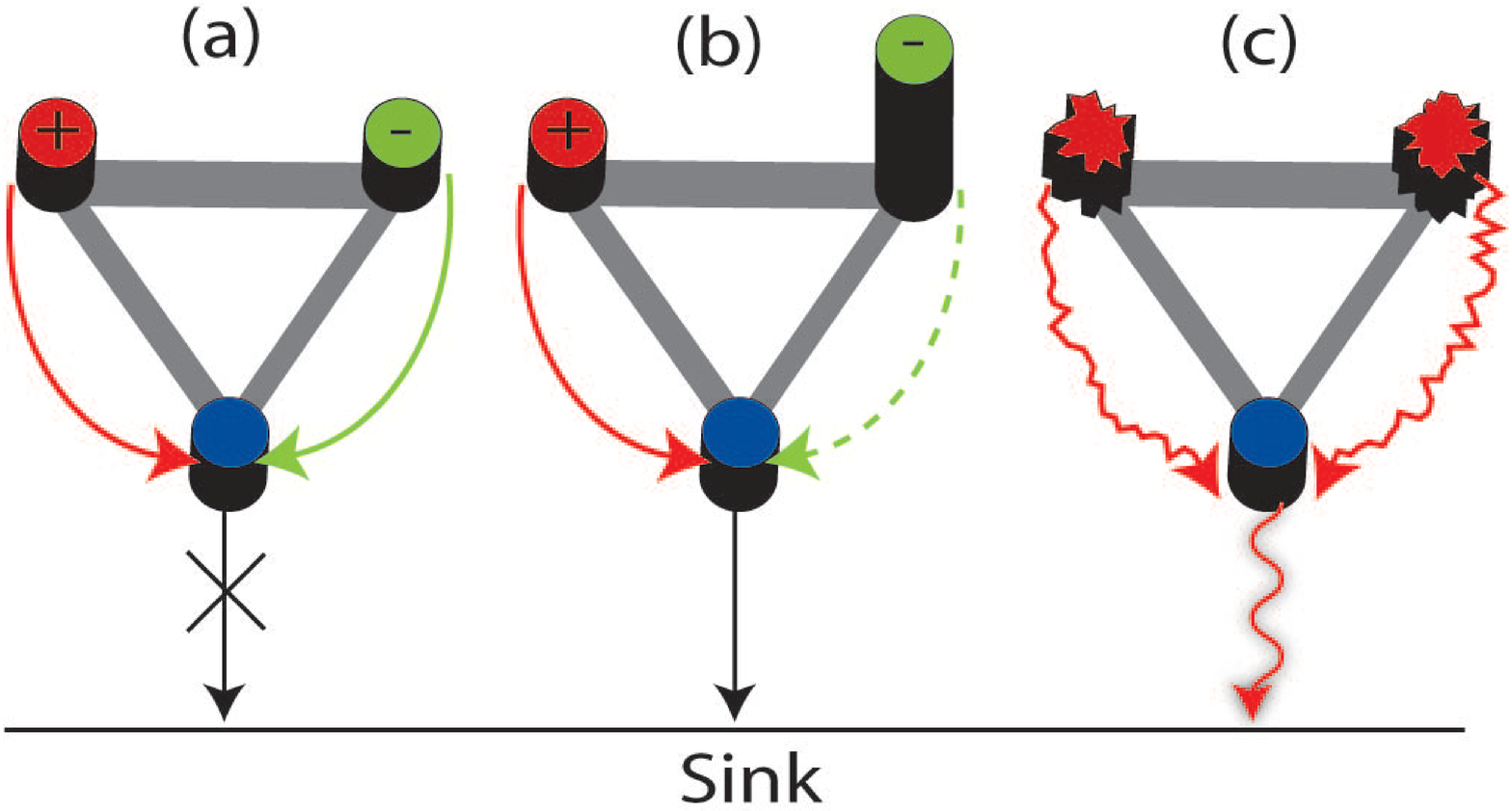}\vspace{-0cm}
\caption{\label{fig:3site} A three-site fully connected network (FCN). In (a) destructive interference of
the tunneling amplitudes from each site in an anti-symmetric
eigenstate prevent transfer to the sink site (Blue). In (b) an
energy mismatch generates an asymmetric stationary state
which is coupled to the sink and leads to transport. In (c), local dephasing efficiently removes destructive interference effects and generally leads to excitation transfer to the sink at a faster time scale than mere energy disorder.}\vspace{-0.5cm}
\end{figure}

We shall show in the next few sections how the combination of
inter-site coherence, interference of tunnelling amplitudes, and
energetic disorder can all conspire to drive strong DAT. For
clarity, we begin by presenting each mechanism separately in simple
network models in which these effects can be isolated. The first
system we will consider is the fully-connected network (FCN). The FCN is
characterized by equal hopping strengths between all sites, i.e.
$\hbar v_{j,l}=J$ for any $j \neq l$, and for the case of a
uniform FCN, i.e. one in which $\omega_{j}$, $\gamma_{j}$, and
$\Gamma_{j}$ are the same on every site, an exact analytical
solution for the density matrix for arbitrarily large networks can
be found~\cite{caro}. We first consider such a uniform FCN with
$\gamma_{j}=0$, $\Gamma_{j}=0$, and one excitation on site $1$.
The only irreversible process left is the decay of population to
the sink from site $N$.
\begin{figure}[t]
\includegraphics[width=5in]{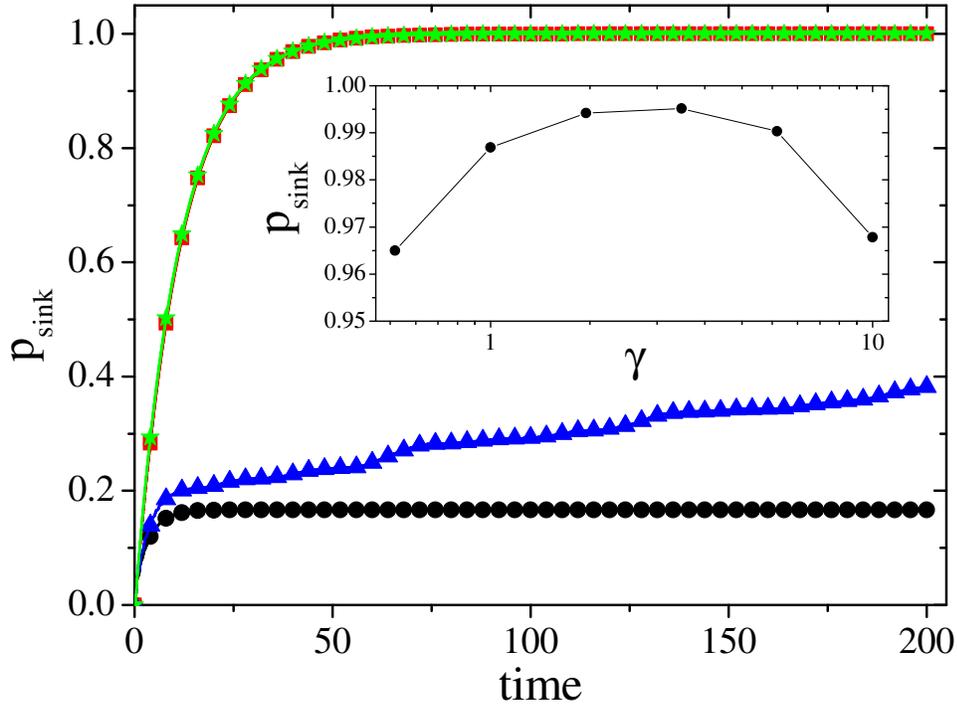}
\caption{\label{fig:psink} $p_{sink}$ vs. time for a FCN of $N=7$ nodes
with $\Gamma=0$, $J=1, \Gamma_{N+1}=1$, for the case of no
dephasing (circles), pure dephasing (squares), static disorder
(triangles), and static disorder with uniform dephasing (star).
Destructive interference in the FCN can also be effectively
removed (in the absence of pure dephasing) by the presence of
static disorder in the local site energies or hopping rates. Such
disorder can prevent the cancellation of tunnelling amplitudes and
thus enhances the asymptotic value of $p_{sink}$. For this case,
the site energies are random numbers drawn uniformly from $[0,1]$,
while the dephasing rates are chosen to equal $1$ for all the
sites in the dephasing case. Inset: $p_{sink}(t)$ at a fixed time
$t=50$ as a function of $\gamma$.}\vspace{-0.5cm}
\end{figure}
In Fig.~(\ref{fig:psink}), the time evolution of the sink population
for these conditions is shown (circles) for the case of $N=7$. This choice is motivated by the fact that the actual FMO complex we will
analyze in subsequent sections can be modelled as 7 sites network.
In the absence of any noisy process, the asymptotic value of $p_{sink}$ is $1/6$ and this should be
contrasted with classical hopping for which $p_{sink}(t\rightarrow
\infty)=1.$ The striking difference between these results can be
seen as a consequence of destructive interference of tunnelling
amplitudes in the quantum case. Although individual sites have
finite amplitudes $J$ for transfer to other sites, a superposition
state of the form
$|\Psi_{ij}\rangle=(|i\rangle-|j\rangle)/\sqrt{2}$ cannot
propagate in the network due to the perfect cancellation of the
tunnelling amplitudes from each state in the antisymmetric
superposition. This coherent trapping is illustrated in Fig.~(\ref{fig:3site}~a). Having identified these non-propagating states, the
excitation asymptotically transferred to the sink can be
understood as being just the weight of the initial state which
lies outside of the `invariant subspace' which consists of all
$|\Psi_{ij}\rangle$ which have zero tunneling matrix elements with the localised state $|N\rangle$ and
therefore do not feel the presence of the sink. For the case
considered here, this insight immediately predicts
$p_{sink}(\infty)=\frac{1}{N-1}$, in agreement with the result
shown in Fig.~(\ref{fig:psink}). Invariant subspace analysis of more
complex networks and other initial conditions can be found in
Ref.~\cite{caro}.

Purely coherent dynamics can thus impede transport in multiply
connected networks via interference, but in the presence of
dephasing noise trapped excitations can become propagating as shown in Fig.~(\ref{fig:3site}~c). This
is the first physical mechanism of DAT: the \emph{removal of
transport-suppressing interference effects}. Fig.~(\ref{fig:psink})
shows that as the dephasing noise strength $\gamma$ increases, the
efficiency rapidly rises to near-perfect excitation transfer as
the phase coherence of the $|\Psi_{ij}\rangle$ states is
destroyed. With further increase in $\gamma$, the efficiency drops
as noise suppresses tunnelling via the quantum Zeno effect, as shown in the figure inset. We
thus find an optimum dephasing strength where the combination of
coherent tunnelling and dephasing maximizes the transport
efficiency, a generic feature of networks which display DAT as we
will discuss again in the context of the FMO dynamics. This
coherent trapping and DAT is related to the type of `dark states' found in
quantum dot and optical systems~\cite{trap2,trap3,trap4}.
If we allow in our model for site energies to become disordered, another
mechanism of DAT appears: \emph{line broadening}. Random energetic
disorder removes the destructive interference discussed above, and
formally leads to $p_{sink}(\infty)=1$ in the absence of noise. Note however the different time scale of the process,
as illustrated by the triangles sequence in Fig.~(\ref{fig:psink}). If
the energy difference between sites is much greater than their
coherent couplings, the energy eigenstates are effectively
localized, and coherent transport between these sites is strongly
suppressed. Thus EET times may be very long and excitations may
decay before reaching the target\cite{caro}. However, pure
dephasing causes spectral broadening of the sites energies, and if
it is sufficiently strong to cause overlap of site energies,
additional incoherent tunnelling between sites can be activated leading
to an increase in EET efficiency as shown in Fig.~(\ref{fig:psink}) (stars).
Line-broadening is also the basis for the Forster mechanism, which
is frequently used to study EET in chromophoric
systems and has been well understood for some time~\cite{forster48}.

\begin{figure}[t]
\includegraphics[width=5in]{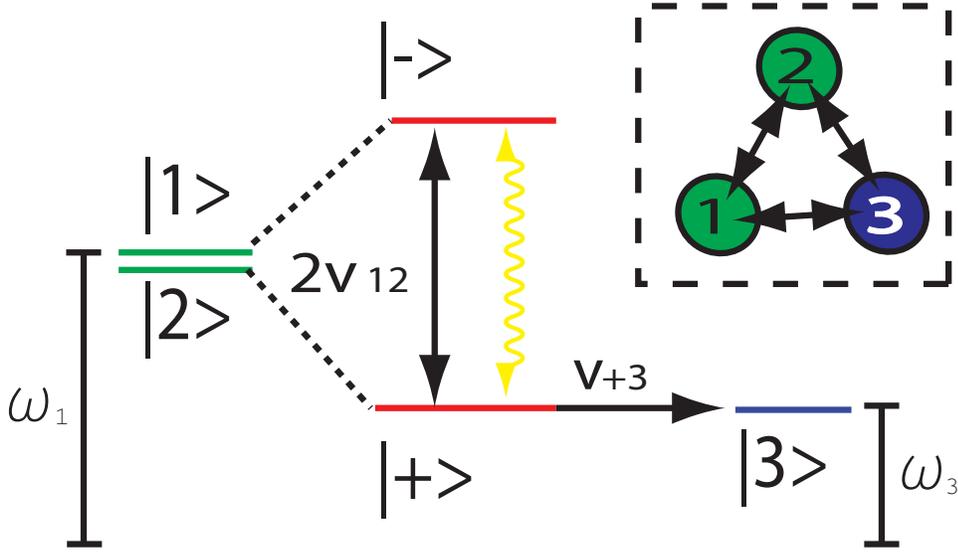}
\caption{\label{fig:hybrid} Energy level and coupling structure in a
hybrid basis for a 3-site FCN (inset). Coherent couplings are
shown by straight arrows. Tuning $v_{12}$ can bring the
$|+\rangle$ state into resonance with site 3 leading to fast
transport. The $|-\rangle$ is uncoupled from site 3, but noise
induces incoherent transitions between $|-\rangle$ and $|+\rangle$
(corrugated arrows), allowing populations initially held in
$|-\rangle$ to also utilize the fast
$|+\rangle\rightarrow|3\rangle$ pathway.} \vspace{-0.5cm}
\end{figure}

It is important to note that highly efficient DAT is not simply
achieved by using noise to drive the dynamics into a purely
classical regime, rather it is the combination of both noise and
coherence, which places the system at the boundary between quantum and
classical physics, that makes DAT much more efficient than purely
coherent dynamics. This interplay and its possible use for optimizing transport efficiencies can be simply illustrated using
a hybridized basis set to look at a slightly non-uniform 3-level
FCN with couplings $v_{13}=v_{23}$ and site energies $\omega_{1}=\omega_{2}\neq\omega_{3}$.
Expressing the FCN Hamiltonian in the basis
$\{|+\rangle,|-\rangle,|3\rangle\}$, where
$\sqrt{2}|\pm\rangle=|1\rangle\pm|2\rangle$, leads to the new level
and coupling structure shown in Fig. (\ref{fig:hybrid}). This figure
shows how the system can exploit coherence to alter the energy
landscape via the coherent splitting of the  $\ket{\pm}$ states, and also to change the hopping matrix elements between states.
In this example, the inter-site coherence can be chosen to create a highly efficient transport pathway to the
sink by bringing the $\ket{+}$ state into resonance with $|3\rangle$ and enhancing the tunnelling amplitude.
However, the other state becomes decoupled from site 3 due to larger energy mismatch and, in this particular case, cancellation of tunnelling matrix elements. Population in this state is decoupled from the sink. Now, adding noise to the system dephases superpositions of localised states $|i\rangle$,
and in the hybrid basis this opens an incoherent transition between $\ket{-}$ and $\ket{+}$. When noise is weak enough to
preserve the coherent level structure for sufficiently long times, the population initially in $\ket{-}$ can therefore also take advantage of the fast resonant $\ket{+}\rightarrow\ket{3}$ transfer leading to a large DAT effect.

\section{Light-harvesting molecules (The FMO Complex)}\label{fmo}

\begin{figure}[t]
\includegraphics[width=5in]{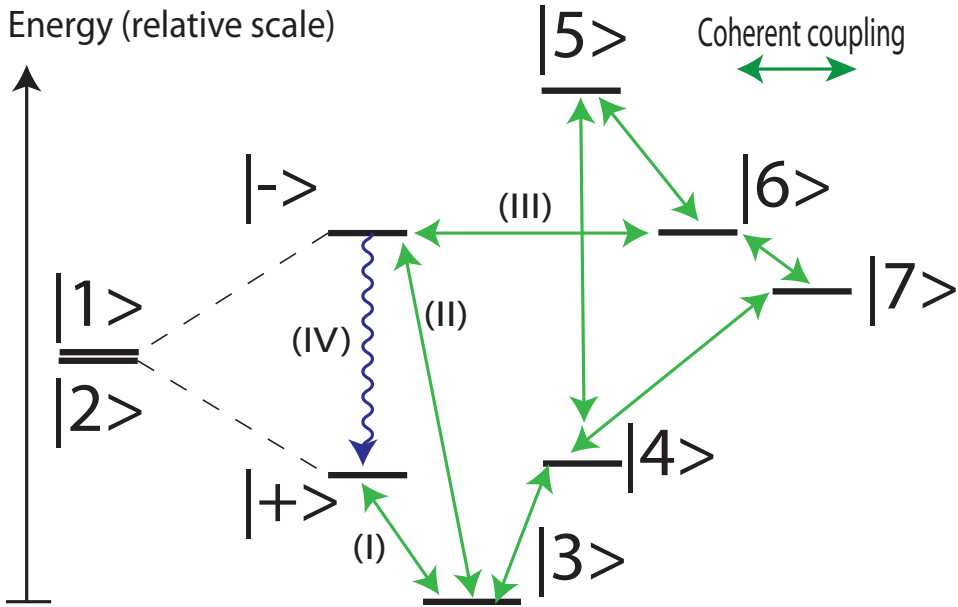}\vspace{-0cm}
\caption{\label{fig:pathways} Energy level structure of the FMO
Hamiltonian in the hybrid basis described in the text. Only the most significant coherent couplings are shown
(green lines), and the coherent interactions which dominate the dynamics in the absence of noise are labeled (I)-(III). For FMO, site $3$ is coupled directly to the sink. The most important new transport pathway arising from the presence of noise is labeled (IV). (a)  Population in $|+\rangle$ decays quickly into the sink due to nearly-resonant coherent interaction with site $3$. (b) Coherent coupling of $|-\rangle$
gives slow population transfer due to large energy mismatch with
site $3$. (c) Resonant coherent coupling causes strong populations
oscillations between $|-\rangle$ and sites $5-7$, inhibiting
transfer to the sink in the absence of noise. This pathway is
suppressed by dephasing and enhances transport. (d) In the hybrid basis, a pure dephasing noise opens an incoherent relaxation channel which allows the population of the $|-\rangle$ state to decay into the $|+\rangle$ state and then quickly decay into the sink via path (a).}\vspace{-0.5cm}
\end{figure}

Having presented some underlying mechanisms of noise-assisted EET
in the FCN models, we now investigate in detail how they operate in the FMO
complex~\cite{olson04}. The FMO complex is a trimer of three
identical units, each composed of seven bacteriochlorophyll $a$
molecules embedded in a scaffolding of protein molecules. We model
the FMO monomer unit as a seven-site network with coupling
strengths and site energies taken from~\cite{adolphs06}. The
typical EET timescale is known to be of several picoseconds, which is
much shorter than the $1$ ns typical radiative lifetime of excitons. The
excitation starts on site 1, thought to be the site closest to the
base plate, and site 3 is connected to the RC (sink). Our
numbering of the FMO sites follows the conventional
one~\cite{olson04,adolphs06}. The excitation transfer in the FMO
complex of \textit{P. aestuarii} undergoing completely coherent
dynamics is shown by the green line in Fig.~(\ref{fig:psinkfmo}), and only reaches $~57\%$
over the typical transfer time of $5$ps. This dephasing-free
evolution resembles the FCN case shown by the sequence of circles in Fig.~(\ref{fig:psink}) and
can be explained using a similar picture.

\begin{figure}[t]
\includegraphics[width=5in]{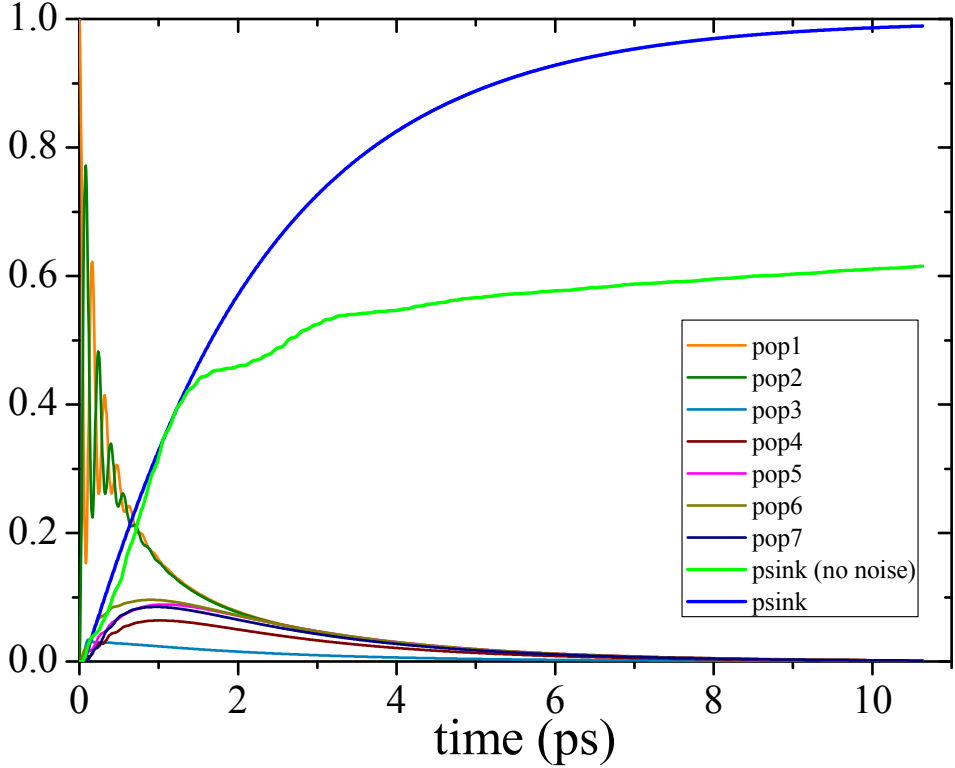}\vspace{-0.4cm}
\caption{\label{fig:psinkmarkov} Site populations vs. time (in $\mathrm{ps}$) for the FMO complex subject to optimized Markovian dephasing noise. We also show $p_{sink}$ for the noiseless case (light green line) and the transfer efficiency for the Linbladian noise model with optimized dephashing rates (blue line). For the optimized decay rates, the transport is nearly complete within the experimental $~5 \mathrm{ps}$ transport time, and coherent oscillations in the dynamics persist until $~1 \mathrm{ps}$. }\vspace{-0.5cm}
\end{figure}

The strong coupling between sites $1$ and $2$ in the FMO
Hamiltonian inspires a hybrid basis
$\{|+\rangle,|-\rangle,|3\rangle...|7\rangle\},$ where $|\pm\rangle=1/\sqrt{2}(|1\rangle\pm|2\rangle)$. In this basis the Hamiltonian has the local site energies and coupling structure shown in Fig.~(\ref{fig:pathways}). An initial excitation on site $1$ corresponds to the initial
condition $\sqrt{2}|1\rangle=|+\rangle+|-\rangle$. As in the FCN
example, the strong coherent interaction between sites $1$ and $2$
pushes the $|+\rangle$ state closer in energy to state $|3\rangle$
and, as shown in Fig. (\ref{fig:pathways}), the population in this state decays quickly into the sink via
path (I). Note however that there is a
relatively small coherent enhancement or cancellation of the
transition amplitudes between $|\pm\rangle$ and $|3\rangle$ as
$v_{13}\ll v_{23}$. The near-resonance of the $|+\rangle$ state
represents the dominant contribution to the initial fast rise in
$p_{sink}$ in the noise-free case as shown in Fig.~(\ref{fig:psinkmarkov}).

Once $\sim 50\%$ of the excitation initially in $|+\rangle$ has
decayed, the rise-time of $p_{sink}$ becomes much slower as the
population initially in $|-\rangle$ does not propagate as efficiently
into site $3$. To understand this we need to examine
the two principal remaining paths (II) and (III) in Fig.~(\ref{fig:pathways}).
The $|-\rangle$ state has roughly the same coupling strength to site $3$ as the $|+\rangle$ state, however it is at a much higher energy. By turning-off all couplings in the Hamiltonian so that only path (II) is active, we are able to measure the isolated population transfer rate via this path, and find that it is over an order of magnitude slower than the decay rate via path
(I) in the absence of noise. The second process (path (III) in Fig.~(\ref{fig:pathways})) would be obscured in a full
site or exciton basis, but reveals itself very clearly in the
hybrid picture. As seen in Fig.~(\ref{fig:pathways}), $|-\rangle$ is almost
resonant with site $6$, and as a result of \emph{constructive}
interference of the tunnelling amplitudes $v_{16}$ and $v_{26}$,
the effective coupling of $|-\rangle$ and $|6\rangle$ is about
twice the energy mismatch of these states.

\begin{figure}[t]
\includegraphics[width=6in]{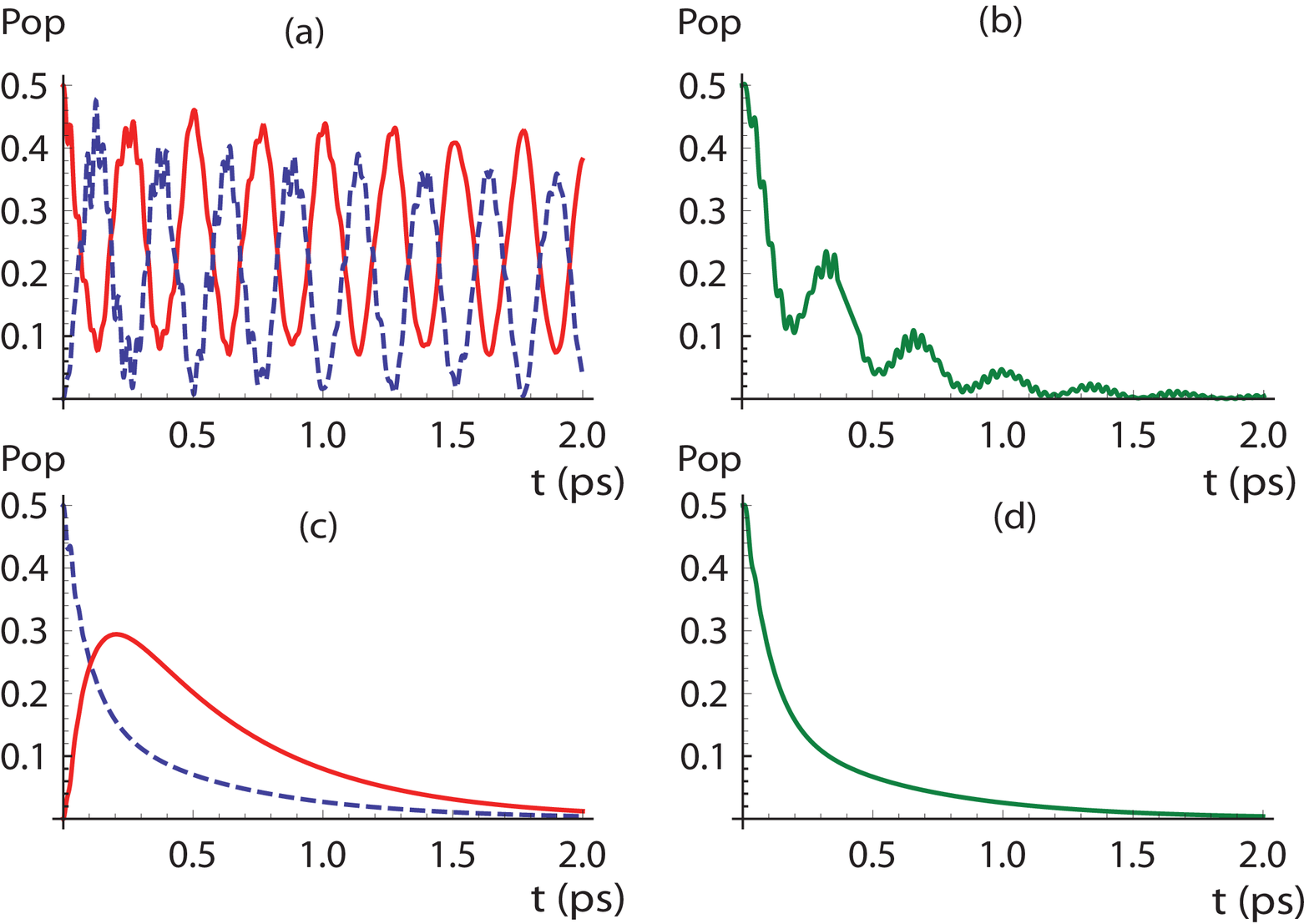}\vspace{-0cm}
\caption{\label{fig:pops} (a) Population of the $|-\rangle$ state (red line) and the combined populations of sites $5-7$ (blue dashed line) as a function of time for the case of no dephasing noise. These populations coherently oscillate due to the resonant coupling of $|-\rangle$ to site $6$. (b) The population of the $|+\rangle$ state in the absence of dephasing. The $|+\rangle$ is close to a direct resonance with site $3$ and this population decays rapidly. This gives the fast initial rise-time of $p_{sink}$ in the absence of dephasing. (c) Population of the $|-\rangle$ state (red line) and the combined populations of sites $5-7$ (blue dashed line) as a function of time with optimized dephasing rates. Coherent oscillations between $|-\rangle$ and $|5-7\rangle$ are suppressed and the population in the $|-\rangle $ states decays rapidly through the incoherent $|+\rangle \rightarrow |3\rangle$ path described in the text. (d) The population of the $|+\rangle$ state with optimized dephasing rates. Oscillatory features are washed out compared to the case of no dephasing, although the average decay rate is similar to case (b). Note also that with the optimized dephasing rates the decay of the $|-\rangle$ state has almost the same time dependence as the $|+\rangle$ state. }\vspace{-0.5cm}
\end{figure}

This strongly-coupled resonance causes the population held in
state $|-\rangle$ to oscillate across the complex between sites $1$ and
$2$ and site $6$. Sites $5$ and $7$, which are strongly coupled to
$6$, also participate in these oscillations and these oscillations are shown in Fig.(\ref{fig:pops})a. Direct and indirect
coherent transport from sites $5-7$ to the sink are even slower
than path (II), and these oscillations along path (III) dominate the dynamics and effectively prevent this population from decaying into the sink via path (II). To
obtain an estimate of the importance of these oscillations we
compare the average rate of change of $p_{sink}(t)$ (after the
initial fast decay) for the full FMO Hamiltonian and the same
Hamiltonian with the coupling between $|-\rangle$ and site $6$ set
to zero. We find that in the latter case the transfer rate becomes $50\%$ larger than the former case.

 The blue line in Fig. (\ref{fig:psinkmarkov}) shows the transport obtained with dephasing rates that were optimized numerically for a transfer time of about $5$ ps~\cite{caro}, and which cause $p_{sink}$ to increase to $0.903$ 
  over this timescale. This dramatic increase in EET arises primarily from the noise-induced suppression of pathway (III) in Fig. (\ref{fig:pathways}) and the new noise-induced incoherent transition between $|-\rangle$ and $|+\rangle$ shown as path (IV) in Fig. (\ref{fig:pathways}).
 Path (IV) allows the population initially in the $|-\rangle$ state to decay via the fast $|+\rangle\rightarrow |3\rangle$ path (I). In fact we find that dephasing on sites $1$ and $2$ alone is sufficient to drive $p_{sink}$ to about $0.85$. The remaining improvement in efficiency in our optimized simulations arises from strong dephasing on sites $3-7$. Dephasing on site $5-7$ (along with dephasing on sites $1$ and $2$) destroys the coherent oscillations that keep the initial $|-\rangle$ population away from the trap in the noiseless case, and the resulting incoherent transitions along path (III) rapidly redistribute this population equally amongst states $|-\rangle$ and $|5\rangle-|7\rangle$. Line broadening due to dephasing on sites $3-7$ quickly transfers populations from sites $5-7$ to the sink via sites $4$ and $3$, and noise on site $3$ further enhances the EET rate via line broadening effects on paths (I) and (II) of Fig. (\ref{fig:pathways}). The suppression of the $|-\rangle\rightarrow |6\rangle$ oscillations and the stronger decay of the $|-\rangle$ population via path (I) of Fig. (\ref{fig:pathways}) are shown in Fig (\ref{fig:pops}c) and Fig. (\ref{fig:pops}d). Most importantly, it is not necessary for the dephasing parameters to be exactly equal to the values obtained by optimization. Large variations in these values still lead to essentially the same evolution of the population in the sink. However, the values used to obtain Fig. (\ref{fig:psinkmarkov}) do give coherent oscillations which last up to about $1$ ps, which is roughly consistent with the experimentally observed coherence time ($>600$ ps) seen in FMO at $77$ K. Similar robustness is also seen to variations in site energies and inter-site couplings. This sort of robustness and effectiveness over a broad range of parameters is essential for the notion of noise assisted transport to operate in natural conditions.

Although there is no direct evidence that natural evolution has performed such a full optimization of the EET process with respect to dephasing noise \cite{lloyd09}, the analysis above provides a good example of how one could in principle optimize EET in an artificial system. By identifying the
naturally effective and inefficient pathways, one could apply noise selectively to create new pathways to circumvent the inefficient ones using the efficient paths as in path (II), enhance the efficient paths with line-broadening as in path (I), or even remove unwanted coherent dynamics such as path (III).

While the microscopic interaction parameters of the FMO pigments and the protein environment are not well characterized experimentally, our local markovian description of the noise neglects a number of potentially important processes, and we now consider two of these. One of these is the role of spatial correlations in the process of EET. The other is the presence of temporal correlations, or memory in the environment that can lead to altered exciton dynamics which are not necessarily governed by a Lindblad type master equation. The first of these, spatially correlated noise can be modelled via the Lindblad term
\begin{equation}
{\cal L}_{deph}(\rho) = - \sum_{mn} \gamma_{mn}  [A_m,[A_n,\rho]],
\end{equation}
where $A_m = \sigma_m^{+}\sigma_m^{-}$. After a full optimization of all $\gamma_{mn}$, we find that $p_{sink}$ saturates at $0.931$ after $5$ ps. This small improvement indicates that non-local effects may be of limited importance for EET in the FMO complex. However, there is experimental evidence for strong spatial correlations in bacterial RC dynamics\cite{noiseRC}, conjugated polymers \cite{collini}, and correlated noise has also been predicted to be relevant in other photosynthetic complexes \cite{olaya}.

\section{Non-Markovian models}

\begin{figure}[t]
\includegraphics[width=5in]{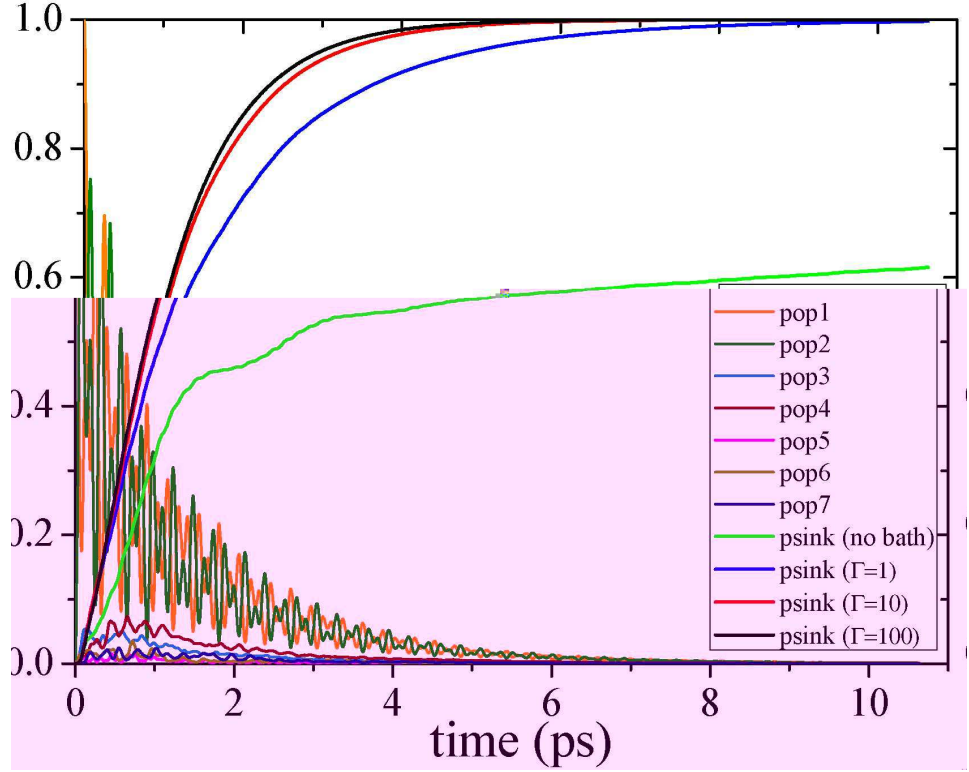}\vspace{-0.4cm}
\caption{\label{fig:psinkfmo} Site population and $p_{sink}$ vs. time (in $\mathrm{ps}$) for the FMO complex. We show the noiseless case (light green line) and the transfer efficiency for a non-Markovian model where each site is coupled to structured phonon bath as specified in ~\cite{adolphs06} and for different values of the local bath damping rate $\Gamma$ expressed in spectroscopic units (cm$^{-1}$). The site population behaviour is plotted for $\Gamma=1$ while oscillations are strongly suppressed for larger values of $\Gamma$.}\vspace{-0.5cm}
\end{figure}

In this last section, we focus on analyzing the possible impact on EET dynamics of deviations from the Markovian model we have presented in the previous sections. As stressed before, the theoretical model used in the previous sections treats pure dephasing
in a way equivalent to the phenomenological Haken-Strobl model. However, a number of recent studies of EET in FMO have looked at the effects of different microscopic models of the exciton-phonon interaction on the dynamics, and in particular at non-markovian dynamics arising from strong coupling and/or the form of the spectral function of the environment \cite{coherence1,guzik,coherence2,thorwart09,prior10}. In this section we shall consider the structured bath spectral density used by Adolphs and Renger in ~\cite{adolphs06}. This spectral density contains a contribution from a low-energy continuous density of states and a discrete high-energy mode, and its effects on the dynamics of a dimer molecule have recently been simulated using a new and numerically exact application of the time-adaptive density matrix renormalization group (t-DMRG) method \cite{prior10}. This study showed that the coupling to the high-energy mode can lead to oscillations in the dimer population dynamics that persist over the whole transport time, and moreover, that the efficiency of inter-site transfer in the presence of these oscillations is just as high as when the high-energy mode is decoupled from the dimer. In order to see if this coexistence of long-lasting oscillations and efficient transport can operate in more complex networks, we now go beyond the dimer setting of Ref.\cite{prior10}, and look at the effects of local mode couplings on the EET dynamics of FMO.

     We will focus here just on the discrete part of the spectral density, and consider a model in which each FMO chromophore is linearly coupled to a harmonic mode with frequency $\omega_H=180 cm^{-1}$, with strength $g=\sqrt{S_H} \omega_h$ and $S_H=0.22$, following~\cite{adolphs06}. To describe these couplings we add to the previous Hamiltonian in Eq. (\ref{ham}) the following two terms
\begin{equation}
H_B = \sum_{j=1}^7 \hbar\omega_h a_j^{+} a_j \; , \\
H_{SB} = \sum_{j=1}^7 g (a_i + a_i^{+}) \sigma_j^{+}\sigma_{j}^{-} \; ,
\end{equation}
with $H_B$ being the free Hamiltonian for the two-level bath with creation and annihilation operators $a^{+}$ and $a$, respectively, and mode frequency $\omega_H$, while $H_{SB}$ is the system-bath interaction Hamiltonian with interaction strength $g$. We note that the frequencies and interactions strength of the couplings predict a reorganization energy, a measure of the coupling strength to the localized mode, of $S_{H}\omega_{H}$, which corresponds to a mode occupation of just $S_{H}=0.22$ when the mode relaxes into equilibrium with respect to the exciton-mode interaction at $T=0$ K on a single isolated chromophore with a single excitation present. It is therefore reasonable to consider the local modes within a two-level approximation, and numerically monitoring of the populations in each local mode confirmed that no local modes were strongly excited or saturated over the whole time interval of the simulations. We also consider the case in which each bath mode can be damped with a damping rate $\Gamma$, which gives a smooth
spectral width to the bath density of states and hence a decay of the correlation time of the environment. When this damping is large, the bath spectral density is essentially unstructured in frequency space and induces a purely Markovian dissipation on the exciton dynamics. The damping is introduced by considering a Lindblad term ${\cal L}^{bath}_{rad}(\rho)$ of the form as in Eq.~(\ref{dissipation}) with the same rate $\Gamma$ for all the local baths.

In Fig.~(\ref{fig:psinkfmo}) we
show the site population behaviour as a function of time for the case of $\Gamma=1$ and the transfer efficiency $p_{sink}$ for different values of $\Gamma$.
Coupling to lightly damped coherent modes ($\Gamma =1$) gives a very large improvement of the transport, as compared with the noiseless case, with $p_{sink} \sim 0.95$
at $5.5~\mathrm{ps}$. This is close to complete transfer.  Note that for this case the populations, notably in sites 1 and 2, display strong oscillations for up to $5.5~\mathrm{ps}$, i.e.,
the whole transport time.  This is perhaps interesting as it shows that persistent oscillations are not necessarily inconsistent with fast and total transport. This should be contrasted to the Lindblad dephasing model of the previous sections, in which we found that excitonic coherences are observable only over the first $~20\%$ of the total transport time when the transport is optmized. Another interesting observation is that this non-markovian model appears to be able to implement the enhancement and suppression of pathways through the FMO energy landscape very well, better in fact than the optimised markov case. Comparing Fig. (\ref{fig:psinkmarkov}) and Fig. (\ref{fig:psinkfmo}) we see that in the local mode simulation, the bath strongly suppresses the transfer of populations from sites $1$ and $2$ to sites $5-7$ which occurs via the resonant path (III) in Fig. (\ref{fig:pathways}) in the absence of noise. While markovian noise evetually destroys these coherent oscillations `across the molecule', there is still a substantial leakage via incoherent tunneling into sites $5-7$and their total population reaches as high as $~0.3$ at $1 \mathrm{ps}$. In the local mode case, the total population of sites $5-7$ never rise above $0.05$. Apart from sites $1$ and $2$ , the only sites to become significantly populated are sites $3$ and $4$, which are closely linked to the sink. The local modes seem to localise the excitations more effectively around the sink and prevents the exploration of inefficient paths like (III) in Fig. (\ref{fig:pathways}).

From the point of view of noise engineering, it seems that the application of non-markovian, quantum noise might allow for a more precise control of the direction and speed of the dynamics. However, it remains to be shown that the strong effect we see is not just due to a fine tuning of exciton and mode parameters in the Hamiltonian, i.e . we would like to see how robust this non-markovian DAT is. This point can be partially investigated by looking at the damping of these modes. As illustrated in Fig. (\ref{fig:psinkfmo}), damping the modes leads to an incremental improvement of the transport efficiency as the damping rate $\Gamma$ acting on each site is increased from the value $\Gamma=1$ (blue curve) to $\Gamma=10$ (red) and $\Gamma=100$ (black). This incremental improvment in transport efficiency coincides with increasingly fast decay of the oscillations of the population dynamics (not shown). This shows that the long-lasting population oscillations in this noise model which are seen for weak mode-damping are likely to involve exciton coherences mediated by the local modes.

Increasing the damping of the local modes makes the effective environment seen by the excitons more markovian in nature and lead to increases in the tranport efficiency as evidenced by the behaviour of $p_{sink}$. This conclusion that increasing markovianity increases the transport rate differs from the one obtained within a different noise model put forward recently in \cite{guzik}, where an enhancement of the transfer rate was observed when the exciton-bath interaction was treated without the markov approximation. It seems clear that linking markovian or non-markovian effects to an improved or slowed down transfer is not straightforward and strongly depends on specific features of the noise model being considered, and perhaps there is no universal, model-independent, relation between non-markovian effects and efficiency of EET. As a result, we can conclude that while the process of EET in FMO is clearly noise-assisted, in the sense that evolution under the exciton Hamiltonian cannot account alone for the observed transport efficiency, the EET dynamics and the transport efficiency as measured by $p_{sink}$ are very sensitive to the specific noise model. In particular, efficient transport with long-lasting oscillations might arise from any combination of weak markovian damping, slow, non-markovian environments as in Refs.\cite{coherence1,thorwart09}, spatially-correlated baths \cite{collini}, or coupling to local modes as shown here.

As the role of noise is clearly crucial for efficient EET, further experimental results are needed to discriminate these different noise models. One important issue related to this is whether or not, at physiological temperatures, the net effect of the dephasing due to the coupling of the complex to the protein environment can be modelled in terms of a classical, fluctuating field or must be treated explicitly as a {\em quantum environment}, like the zero temperature localized mode model discussed in this section. The dynamical behaviour of quantum correlations amongst the excitons \cite{coherence2}, and also between the bath and exciton system \cite{taka}, is also expected to be sensitive to the noise model, and the recent formulation of efficient techniques for the tomographic characterization of many-body systems \cite{tomog} may allow for experiments which can directly probe the nature of the exciton-protein coupling in photosynthetic complexes.

\section{Conclusions}

We have revisited the basic mechanisms by which pure dephasing noise can open up or suppress pathways through the energy landscape of a quantum network, and have shown how these processes can dramatically increase the eficiency of EET relative to their noiseless evolution. Using these insights, we contructed a hybrid basis for analysing the FMO dynamics and indicated the key pathways which are responsible for the inefficiency of the EET in the absence of noise. For excitations to propagate efficienctly, these pathways must either be avoided or inhibited, and we have shown within a fully Markovian approach how pure dephasing noise acheives this, and how it effects each of the principal pathsways individually. As the identification of the transport-suppressing pathways is made without any reference to the noise model, we believe that any noise model that enhances tranport must more or less carry out this strategy of supresing the inefficient paths and enhancing the direct relaxation channels. This intuition is supported by our results using the non-markovian local mode model, where we find a very effective suppression of the inefficient pathways, leading to a transport time which can even exceed the optimised markovian dynamics. The results obtained from these models emphasize how knowledge of the underlying Hamiltonian, when expressed in an appropriate basis, might allow the possiblility of using appplied noise almost as an enginering tool for the creation of artifical light-harvsting architectures. Though we have demonstrated the technique for the Fenna-Matthew-Olson (FMO) complex, we could use such a path analyisis to quite literally follow the movement of the excitation across any network to the sink, and enhance the energy flow with selective application of appropriate noise interactions.

We have also shown in our local mode noise model, that the experimentally observed transfer time can occur through dynamics which preserve coherences and entanglement across the whole transport time. The optmized markovian theory can only preserve coherences for $~20\%$ of this time. The potential relevance of this observation for understanding the long-lasting coherences observed in FMO and other photosynthetic complexes still remains unclear, and in particular, the effects of temperature must be understood before a connection to the experimental data can be made. However, this model provides an interesting new system for studying \textit{quantum noise}-assisted transport, and further emphaises one of the underlying message of this paper, which is; the neccesary tasks of enhancing exciton transport revealed by our path analysis can be acheived by a rich variety of noise models, each of which can generate very different dynamics. As noise is a crucial part of the efficiency of EET in photosynthetic complexes, the need for experimental discrimination of noise models is paramount. Such studies would also provide clues to the still unanswered question of whether quantum mechanics, in the form of coherence and entanglement, is necessary for EET or if it is just an inevitable consequence
of quantum mechanical evolution on the short length and timescales found in pigment-protein complexes. Answering this question will require further studies in energy transport across connected networks in the presence
of various noise models, with interesting ramifications for the role of coherence and entanglement in the dynamics of interacting systems.

\ack 
This work was supported by the EPSRC QIP-IRC, EPSRC grant EP/C546237/1, EU projects QAP and CORNER, the von Humboldt Foundation and the Royal Society. We are grateful to G.~R. Fleming, A. Ishizaki, S. Virmani and T. Brandes for helpful discussions and comments. F.~C. was supported also by a Marie Curie Intra European Fellowship within the 7$^{\rm th}$ European Community Framework Programme.

\end{document}